\documentstyle[12pt,a4]{article}

\begin{document}

\renewcommand{\thefootnote}{*}
\begin{center}
{\Large\bf The structure of the quantum mechanical state space and induced
superselection rules} \footnote
{Invited lecture at the Workshop on Foundations of Quantum Theory, T.I.F.R.
Mumbay,\protect\\ September 9. -- 12., 1996} 
\\[2cm]
{\large\bf Joachim Kupsch} \\
{\large Theoretical Physics Group\\
Tata Institute of Fundamental Research\\
Mumbay 400 005, India\\[0.5cm]
and\\[0.5cm]
Fachbereich Physik, Univ. Kaiserslautern\\
D-67653 Kaiserslautern, Germany 
\renewcommand{\thefootnote}{\dagger}  \footnote
{Permanent address. e-mail: kupsch@physik.uni-kl.de} \\[0.5cm]
November 1996}\\[1.5cm]
\end{center}
\begin{abstract}
The role of superselection rules for the derivation of classical probability
within quantum mechanics is investigated and examples of superselection
rules induced by the environment are discussed.
\end{abstract}

\section{Introduction}

One of the puzzles of quantum mechanics is the question, how classical
objects can arise in quantum theory. Quantum mechanics is a statistical
theory, but its statistics differs on a fundamental level from the
statistics of classical objects. The EPR problem and the violation of Bell's
inequalities are consequences of this fact [1--3](English translation of [2]
in [4]). These questions are usually discussed as problems of the
interpretation of the Schr{\"o}dinger wave function. Here I would like to
emphasize a more abstract approach which relates these problems to the
geometrical structure of the quantum mechanical state space, i.e. the total
set of pure and mixed states. It is exactly this structure which forbids to
assign an objective probability measure to the states.

It is known since a long time that the statistical results of quantum
mechanics become consistent with a classical statistics of ``facts'', if the
superposition principle is reduced to ``superselection sectors'', i.e.
coherent orthogonal subspaces of the full Hilbert space. The mathematical
structure of quantum mechanics and of quantum field theory provides us with
only a few ``superselection rules'', the most important being the charge
superselection rule related to gauge invariance, see e.g. [5,6] and the
references given therein. But there are definitively not enough of these
superselection rules to understand classical properties in quantum theory. A
possible solution of this problem is the emergence of effective
superselection rules due to decoherence caused by the interaction with the
environment. These investigations -- often related to a discussion of the
process of measurement -- started in the eighties; some early references are
[7--9].

In this talk the transition from quantum mechanics to classical physics is
investigated in the following restricted sense: How can a classical
statistical theory emerge from quantum probability? The emphasis is laid on
the principles of induced superselection. Many aspects, like Markov
approximation and localization models are totally omitted. For a
comprehensive investigation of decoherence see the book [10]. The talk is
organized as follows. The structure of the state space is discussed in
Sect.2. Superselection rules are defined in Sect.3. In Sect.4 the dynamics
of subsystems and the emergence of effective superselection rules are
recapitulated. Some exactly solvable models and mathematical formulations of
induced superselection rules are discussed in Sect.5.

\section{Structure of the state space\label{ssp}}

We start with a few mathematical notations. If ${\cal H}$ is a separable
Hilbert space we use the following spaces of linear operators defined on $%
{\cal H}$.

${\cal B}({\cal H})$: The {\bf R}-linear space of all bounded selfadjoint
operators $A$. The norm of this space is the operator norm $\Vert A\Vert$.

${\cal T}({\cal H})$: The {\bf R}-linear space of all selfadjoint traceclass
operators $A$. These operators have a pure point spectrum ${\alpha_i}\in 
{\bf R}$, i=1,2,..., with $\sum_i\vert\alpha_i\vert < \infty$. The natural
norm of this space is the trace norm $\Vert A\Vert_1={\rm tr}\sqrt{A^{+}A}%
=\sum_i\vert\alpha_i\vert$. Another norm, used in the following sections, is
the Hilbert-Schmidt norm $\Vert A\Vert_2=\sqrt{{\rm tr}A^{+}A}$. These norms
satisfy the inequalities $\Vert A\Vert\leq\Vert A\Vert_2\leq\Vert A\Vert_1$.

${\cal D}({\cal H})$: The set of all statistical operators, i.e. positive
traceclass operators $W$ with a normalized trace, ${\rm tr}W = 1$.

${\cal P}({\cal H})$: The set of all rank one projection operators $P^1$.

\noindent These sets satisfy the obvious inclusions ${\cal P}({\cal H}%
)\subset {\cal D}({\cal H})\subset {\cal T}({\cal H})\subset {\cal B}({\cal H%
}).$

Any state of a quantum system is represented by a statistical operator $W \in%
{\cal D}({\cal H})$, the elements of ${\cal P}({\cal H})$ thereby correspond
to the pure states. Any observable is represented by an operator $A\in {\cal %
B}({\cal H})$. The use of bounded operators is only an apparent restriction,
since the whole information of an unbounded observable can be recovered from
its spectral resolution involving only bounded operators. The expectation of
the observable $A$ in the state $W$ is the usual trace ${\rm tr}WA$, which
is always finite for $W\in{\cal D}({\cal H})$ and $A\in{\cal B}({\cal H})$.
Without additional knowledge about the structure of the system we have to
assume that the set of all states corresponds exactly to ${\cal D}({\cal H})$%
, and the set of all (bounded) observables to ${\cal B}({\cal H})$. The
state space ${\cal D}({\cal H})$ has an essential property: it is a convex
set, i.e. $W_1, W_2\in {\cal D}({\cal H})$ implies $\lambda_1W_1+%
\lambda_2W_2\in {\cal D}({\cal H})$ if $\lambda_{1,2}\geq 0$ and $%
\lambda_1+\lambda_2=1.$ Any statistical operator $W\in {\cal D}({\cal H})$
can be decomposed into pure states 
\begin{equation}
\label{5}W=\sum_n{\lambda}_nP_n^1 
\end{equation}
with $P_n^1\in {\cal P}({\cal H})$ and probabilities $w_n\geq 0$, $%
\sum_nw_n=1$. An explicit example is the spectral decomposition of $W$. But
there are many other possibilities, and we shall investigate that aspect in
more detail. It is exactly this arbitrariness that does not allow a
classical interpretation of quantum probability.

Before we continue with the investigation of the quantum mechanical state
space we introduce some notations concerning convex sets. A bounded and
closed subset ${\cal M}$ of an {\bf R}-linear space is convex if with two
points $x_1, x_2 \in {\cal M}$ also the connecting line segment ${\lambda}%
x_1+(1-{\lambda})x_2, {\lambda}\in [0,1]$, belongs to ${\cal M}$. Those
points of ${\cal M}$ which do not lie on a line connecting two others, i.e.
for which $x={\frac{1}{2}}x_1+{\frac{1}{2}}x_2$ with $x_{1,2}\in {\cal M}$
is only possible if $x_1=x_2=x$, are called extremal. The extremal points
are always boundary points, but not necessarily all boundary points are
extremal. The boundary of ${\cal M}$ will be denoted by $\partial{\cal M}$,
the set of extremal points -- the {\it extremal boundary} -- by $\partial_e%
{\cal M}$. An important statement about convex sets is: Any point $x \in 
{\cal M}$ can be represented by an integral over the extremal boundary 
\begin{equation}
\label{6}x=\int_{\partial_e{\cal M}}y d\lambda(y) 
\end{equation}
where $d\lambda(y)$ is a probability measure, i.e. a non-negative measure
with $\int d\lambda(y)=1$, concentrated on (the closure of) $\partial_e{\cal %
M}$. This representation has been derived in a rather general context by
Choquet, see e.g. [11,12]. In the case of point measures the integral (\ref
{6}) yields the sum 
\begin{equation}
\label{7}x=\sum_i\lambda_iy_i 
\end{equation}
with $y_i\in \partial_e{\cal M}$ and probabilities $0\leq \lambda_i \leq 1,
\sum_i\lambda_i=1.$ The representations (\ref{6}) and(\ref{7}) are only
unique if ${\cal M}$ is a simplex. An n-dimensional simplex ${\cal M}$ is
the closed convex set generated by n+1 points $y_i, i=1,...,n+1,$ where the
connecting lines $y_i-y_{n+1}, i=1,...,n$ are linearly independent, ${\cal M}%
=$\{$x=\sum_{i=1,...,n}\lambda_iy_i \vert \lambda_i\geq 0,\sum_i\lambda_i=1$%
\}. In two dimensions this is a triangle, in three dimensions a tetrahedron.
The extremal boundary of a simplex is the set of its vertices, and the
weights $\lambda_i$ are the uniquely defined barycentric coordinates of the
point $x\in {\cal M}$ with respect to the vertices $y_i$. Another class of
convex sets are balls $B^n=\{x\in {\bf R}^n , \vert x \vert\leq1\}$. It is
easily seen that in this case the extreme boundary coincides with the
boundary $S^{n-1}=\{x\in{\bf R}^n , \vert x \vert=1\}$. For balls the
representations (\ref{6}) or (\ref{7}) are highly non-unique, since any $%
y\in \partial_e{\cal M}$ can show up on the right hand side with positive
weight.

The essential difference between the state space of classical mechanics and
that of quantum mechanics is:

\noindent -- The state space of classical mechanics is an (infinite
dimensional) simplex with a unique representation (\ref{6}).

\noindent -- The state space of quantum mechanics is not a simplex, its
structure is closer to that of a ball. The decompositions (\ref{6}) or (\ref
{7}) into extremal elements, i.e. pure states as explained below, are highly
non-unique.

The state space of classical mechanics is the set of all probability
measures $d\mu(\xi)$ on the phase space $\Xi\subset {\bf R}^n$ of a given
system. The extremal boundary of that set is given by all Dirac measures $%
\{\delta_{\eta}(\xi)d^n\xi=\delta(\xi-\eta)d^n\xi \vert \eta\in\Xi\}$. The
representation (\ref{6}) is the identity $d\mu(\xi)=\int\delta_{\eta}(\xi)d%
\mu(\eta)$, which leads again to the originally given measure. For the
mathematical proof that the space of probability measures is an infinite
dimensional simplex with a unique Choquet representation see [11,12].

We have already seen that the state space of quantum mechanics is convex.
The boundary of this set is formed by those statistical operators which have
at least one eigenvalue zero. The condition for extremal elements, i.e. $W={%
\frac{1}{2}}W_1={\frac{1}{2}}W_2$ has to imply $W_1=W_2$, is met by all rank
one projection operators $W\in{\cal P}({\cal H})$. The decomposition (\ref{5}%
) is therefore the Choquet representation (\ref{7}) of an element of a
convex set. As a consequence of the superposition principle this
representation is highly non-unique, see e.g. Sect.2.3 of [13]. As explicit
example we consider the state space of a spin-${\frac{1}{2}}$ system, i.e. $%
{\cal D}({\bf C}^2)$. Since all statistical operators on ${\bf C}^2$ with
one eigenvalue zero are rank one projectors, the boundary of the state space
coincides with the extreme boundary. More explicitly, any operator $\rho\in 
{\cal D}({\bf C}^2)$ can be represented with a real polarization vector $%
\overrightarrow{p}$ in the unit ball $B^3=\{\overrightarrow{p}\in {\bf R}^3
\vert \vert\overrightarrow{p}\vert \leq 1\}$ as 
\begin{equation}
\label{8}\widehat{\rho }(\overrightarrow{p}):=\frac 12\left( {\bf 1}+%
\overrightarrow{\sigma }\overrightarrow{p}\right) 
\end{equation}
where ${\bf 1}$ is the unit 2x2 matrix and $\overrightarrow{\sigma}%
=(\sigma_1,\sigma_2,\sigma_3)$ are the Pauli matices. Moreover, a simple
calculation leads to the identity $\Vert\rho(\overrightarrow{p}_1)-\rho(%
\overrightarrow{p}_2)\Vert_1=\vert \overrightarrow{p}_1-\overrightarrow{p}_2
\vert$. Hence the topology of ${\cal D}({\bf C}^2)$ given by the trace norm
agrees with the euclidean metric of the polarization vectors. The set ${\cal %
D}({\bf C}^2)$ is therefore isomorphic to the unit ball $B^3$. The extremal
elements corresponding to $\overrightarrow{p}\in S^2$, are exactly the
projection operators of the representation (\ref{5}). Since $\lambda_1\rho(%
\overrightarrow{p}_1)+\lambda_2\rho(\overrightarrow{p}_2)=\rho(\lambda_1%
\overrightarrow{p}_1+\lambda_2\overrightarrow{p}_2)$ if $\lambda_{1,2}\geq0$%
, $\lambda_1+\lambda_2=1$, the representation (\ref{5}) corresponds
therefore to the Choquet representation (\ref{7}) of the unit ball $B^3$.
The non--uniqueness of that representation is therefore obvious.

In higher dimensions the geometrical picture of ${\cal D}({\cal H})$, dim$%
{\cal H}\geq 3$, is more complicated. The boundary is much larger than the
extremal boundary, but the arbitrariness of the representation (\ref{5})
remains. To see this one can choose any two dimensional subspace of the
spectral decomposition of $W$ with at least one non-zero eigenvalue. Then
this part of the spectral representation corresponds up to normalization to $%
{\cal D}({\bf C}^2)$, and we can modify it with all the arbitrariness seen
above.

\section{Superselection rules\label{ssr}}

The arbitrariness of the decomposition (\ref{5}) originates in the
superposition principle. In quantum mechanics, especially in quantum field
theory, the superposition principle can be restricted by superselection
rules. Here we cannot discuss the arguments to establish such rules, for
that purpose see e.g. [5,6], and also Chap.6 of [10], or to refute them, see
e.g. [14]. Here we only investigate the consequences for the structure of
the state space. In a theory with superselection rules like the charge
superselection rule, the Hilbert space ${\cal H}$ splits into orthogonal
superselection sectors ${\cal H}_m,m\in {\bf M,}$ such that ${\cal H=}\oplus
_m{\cal H}_m$. Pure states with charge $m$ (in appropriate normalization)
are then represented by vectors in ${\cal H}_m$, and superpositions of
vectors with different charges have no physical interpretation. The
projection operators $P_m$ onto the orthogonal subspaces ${\cal H}_m$
satisfy $P_m P_n=\delta_{mn}$ and $\sum_mP_m =I$. The set of states is
reduced to those statistical operators which satisfy $P_mW=WP_m$ for all
projection operators $P_m,m\in {\bf M}$. The state space of the system is
then ${\cal D}^S=\{W\in{\cal D}({\cal H})\vert WP_m=P_mW, m\in {\bf M}\}$,
and all statistical operators satisfy the identity 
\begin{equation}
\label{9}W=\sum_mP_mWP_m. 
\end{equation}
A mathematical equivalent statement is that all observables of such a theory
commute with the projection operators $P_m,m\in {\bf M}$. Superselection
rules of this type will be called ``kinematical superselection rules'' to
contrast them against the ``dynamically induced superselection rules'' to be
discussed in the following section.

As a consequence of (\ref{9}) the extremal boundary of ${\cal D}^S$
decomposes into ${\cal P}^S= \cup_m {\cal P}^S_m$ with ${\cal P}^S_m=\{P^1\in%
{\cal P}({\cal H})\vert P^1{\cal H}\subset {\cal H}_m\}$. The decomposition
of a statistical operator $W\in{\cal D}^S$ into pure states now reads (here
the sum over i can be substituted by an integral over the set ${\cal P}^S_m$%
) $W=\sum_{m,i}\lambda_{m,i}P_{m,i}^1$ with $P_{m,i}^1\in{\cal P}^S_m$ and
probabilities $\lambda_{m,i}\geq0$, $\sum_{m,i}\lambda_{m,i}=1$. Since the
representation of an element of ${\cal D}({\cal H}_m)$ by pure states is not
uniquely given (if dim${\cal H}_m\geq2$), the right hand side of of this
decomposition is again highly non-unique. But as a consequence of the
structure of the state space ${\cal D}^S$ the total probability with respect
to any of the subsets ${\cal P}^S_m$, i.e. $\sum_i\lambda_{m,i}$, has no
ambiguity since $\sum_i\lambda_{m,i}={\rm tr}WP_m$, where the right hand
side is uniquely defined. Hence superselection rules allow to speak about
objective ``properties'' of a quantum system. These properties show up with
the probabilities tr$WP_m$ irrespective of what other specifications
(measurements) are made. In the language of observables these properties are
represented by the selfadjoint operators $P_m$, which commute with all
observables of the system. The importance of superselection rules for the
transition from quantum probability to classical probability is obvious. But
there remains an essential problem: Only very few superselection rules can
be found in quantum mechanics that are compatible with the mathematical
structure and with experiment. A satisfactory solution to this problem is
the emergence of effective superselection rules induced by the interaction
with the environment.

\section{Dynamics of subsystems\label{subsys}}

In the following we consider an ``open system'', i.e. a system $S$ which
interacts with an ``environment'' $E$, such that the total system $S+E$
satisfies the usual Hamiltonian dynamics. The system $S$ is singled out by
the fact that all observations refer only to this subsystem. The Hilbert
space ${\cal H}_{S+E}$ of the total system $S+E$ is the tensor space ${\cal H%
}_S\otimes {\cal H}_E$ of the Hilbert spaces for $S$ and for $E $. We assume
that the only observables at our disposal are the operators $A\otimes I_E$
where $A\in {\cal B}({\cal H}_S)$ is an arbitrary bounded selfadjoint
operator on ${\cal H}_S$. If the state of the total system is $W\in {\cal D}(%
{\cal H}_{S+E})$, then all expectation values ${\rm tr}W(A\otimes I_E)$ can
be calculated from the reduced statistical operator $\rho ={\rm tr}_EW$
which is an element of ${\cal D}({\cal H}_S)$ defined such that the
expectation values satisfy ${\rm tr}_S\rho A={\rm tr}_{S+E}W(A\otimes I_E)$.
Since all information about a physical subsystem is given by a statistical
operator, we shall here refer to the statistical operator as the ``state''
of the subsystem. (The state of the total system cannot be recovered from
these ``states'' of its subsystems.)

As mentioned above we assume the usual Hamiltonian dynamics for the total
system, i.e. $W(t)=U(t)W(0)U^{+}(t)$ with the unitary group $U(t)$,
generated by the total Hamiltonian. Except for the trivial case that $S$ and 
$E$ do not interact, the dynamics of the reduced statistical operator $%
\rho(t)={\rm tr}_EU(t)W(0)U^{+}(t) $ is no longer unitary. It is the purpose
of the second part of this talk to evaluate this dynamics in some detail.
The essential result is that this dynamics can produce effective
superselection sectors, i.e. 
\begin{equation}
\label{12}\rho (t)\cong \sum_mP_m\rho (t)P_m 
\end{equation}
in sufficiently short time with a set of projection operators $P_m,m\in {\bf %
M}$, which correspond to a superselection structure (\ref{9}). In Sect.\ref
{models} we shall give more precise formulations of (\ref{12}). The
suppression of the off-diagonal terms of the statistical operator in (\ref
{12}) is essential for the emergence of classical properties in quantum
mechanics. This suppression can never be understood by semiclassical
approximations alone.

\section{Solvable models\label{models}}

The statement (\ref{12}) is so far rather vague since it does not specify
the asymptotics. The following examples show what type of asymptotics is
possible in principle. For a restricted class of models an estimate 
\begin{equation}
\label{13}\Vert \rho (t)-\sum_nP_n\rho (t)P_n\Vert _2=\left\| \sum_{m\neq
n}P_m\rho (t)P_n\right\| _2\leq C_\gamma (1+\delta \left| t\right|
)^{-\gamma } 
\end{equation}
with the Hilbert-Schmidt norm is possible for arbitrary $\rho (0)\equiv \rho
\in {\cal D}({\cal H}_S)$. The constants $\gamma >0,$ $\delta >0$ and $%
C_\gamma >0$ do not depend on $\rho $. Since one can achieve large values of 
$\gamma $ and/or small values of the constant $C_\gamma ,$ these dynamically
induced superselection sectors $P_n{\cal H}$ cannot be distinguished
practically from the kinematical superselection sectors (\ref{9}). For a
subclass of these models one can even derive an estimate of the type (\ref
{13}) with the stronger trace norm. A consequence of (\ref{13}) is that the
transition between a superselection sector $P_n{\cal H}$ and its complement $%
\widehat{P}_n{\cal H}:=(I-P_n){\cal H}$ is suppressed by $\left\| \widehat{P}%
_n\rho (t)P_n\right\| _2\leq C_\gamma(1+\left| t\right| )^{-\gamma }$ again
with the Hilbert-Schmidt norm. If $P$ is a projection operator with finite
rank $N$ on a subspace of $P_n{\cal H,}$ i.e. $P_nP=PP_n=P,$ then $\widehat{P%
}_n\rho (t)P$ is a traceclass operator with $\left\| \widehat{P}_n\rho
(t)P\right\| _1\leq \sqrt{N}C_\gamma (1+\left| t\right| )^{-\gamma },$ and
the transition between $P{\cal H}$ and all other sectors $P_m{\cal H}$, $%
m\neq n,$ is uniformly suppressed in trace norm. The investigation of the
models also shows that unfortunately such simple estimates are rather
unstable against slight modifications of the models.

The models have the following structure. The Hilbert space is ${\cal H}%
_{S+E}={\cal H}_S\otimes {\cal H}_E.$ The total Hamiltonian has the form 
\begin{equation}
\label{14a}H=H_S\otimes I_E+I_S\otimes H_E+V_S\otimes V_E+V 
\end{equation}
where $H_S$ is the Hamiltonian of S, $H_E$ is the Hamiltonian of E, $%
V_S\otimes V_E$ is the interaction term between S and E with selfadjoint
operators $V_S$ on ${\cal H}_S$ and $V_E$ on ${\cal H}_E$, and $V$ is a
possible additional scattering potential. For all models we assume that

\medskip
\noindent 1) The operators $H_E$ and $V_E$ commute, $\left[ H_E,V_E\right]
=O $, hence $\left[I_S\otimes H_E,V_S\otimes V_E\right] =O.$

\noindent 2) The operator $V_S$ has a pure point spectrum 
\begin{equation}
\label{15}V_S=\sum_m\lambda _mP_m 
\end{equation}
where the projection operators $P_m$ form a complete set, $P_mP_n=\delta
_{m,n}P_n,$ and $\sum_mP_m=I.$ We assume that the eigenvalues $\lambda _m$
are separated with the lower bound $\left| \lambda _m-\lambda _n\right| \geq
\delta >0$ if $m\neq n.$

\noindent 3) The operator $V_E$ has an (absolutely) continuous spectrum.

\medskip
Remark. The continuous spectrum of $V_E$ is needed to obtain simple
estimates for $t\rightarrow \infty $. But one could also allow an operator
with point spectrum (as done in [8]), if the spacing of the eigenvalues is
sufficiently small. Then the norm in (\ref{13}) is an almost periodic
function and the inequality is only correct for a finite time interval $%
0\leq t\leq T$. But $T$ can be large enough for all practical purposes.

\subsection{The Araki-Zurek model\label{mod1}}

The first solvable models to discuss the reduced dynamics have been given by
Araki [7] and Zurek [8], and the following construction is essentially based
on these papers. In addition to the specifications made above, we demand: 
\medskip

\noindent4) The operators $H_S$ and $V_S$ commute, the potential $V$
vanishes, i.e. $\left[ H_S,V_S\right]=O=V$. \medskip

\noindent For an originally factorizing state $W=\rho \otimes \omega $ we
calculate with $U(t) =$ exp$(-iHt)$ 
\begin{equation}
\label{16}\rho (t)={\rm tr}_EU(t)WU^{+}(t)={\rm e}^{-\imath
H_St}\sum_{m,n}P_m\rho P_n{\rm e}^{\imath H_St}\chi _{m,n}(t) 
\end{equation}
with $\chi _{m,n}(t)={\rm tr}\left( {\rm e}^{-\imath (\lambda _m-\lambda
_n)V_Et}\omega \right),$ see e.g. Sect.7.6 of [10]. The trace $\chi
_{m,n}(t) $ vanishes for $|t|\rightarrow \infty $ if $m\neq n$ since $V_E$
has an absolutely continuous spectrum. Under additional smoothness
assumptions on $\omega $ (which do not restrict $\rho \in {\cal D}({\cal H}%
_S)$) we derive for $\chi (t):={\rm tr}\left( {\rm e}^{-\imath V_Et}\omega
\right) $ the estimate $\left| \chi (t)\right| \leq C_\gamma (1+\left|
t\right|)^{-\gamma }.$ Here $\gamma $ can be arbitrarily large if $\omega $
is a sufficiently differentiable function in the spectral representation of $%
V_E$ (and vanishes at the boundary points of the spectrum). This estimate
leads to the upper bound 
\begin{equation}
\label{17}|\chi _{m,n}(t)|\leq C_\gamma (1+\delta \left| t\right| )^{-\gamma
} 
\end{equation}
if $\left| \lambda _m-\lambda _n\right| \geq \delta >0$, and we obtain the
estimate (\ref{13}). If we have only a finite number of eigenvalues in (\ref
{15}), a bound of the type (\ref{13}) with the stronger trace norm can be
derived.

This result depends on the reference state $\omega $ only via the decrease
of $\chi(t)$. We could have chosen a more general initial state $W=\sum_\mu
\rho _\mu \otimes \omega _\mu \in {\cal D(H}_{S+E})$ with $\rho _\mu \in 
{\cal T(H}_S{\cal )}$, $\omega _\mu \in {\cal D}({\cal H}_E)$. The operators 
$\rho _\mu $ need not to be positive separately, but we have of course $\rho
=\sum_\mu \rho _\mu \in {\cal D(H}_S{\cal ).}$ Then (\ref{13}) is still
valid if $\sum_\mu \left| {\rm tr}\left( {\rm e}^{-\imath (\lambda
_m-\lambda _n)V_Et}\omega _\mu \right) \right| $ satisfies a uniform
estimate (\ref{17}). Hence the emergence of the superselection sectors $P_n%
{\cal H}_S$ is not sensitive to the initial conditions.

\subsection{A spin model\label{mod2}}

So far we have made the rather restrictive assumption $V=\left[
H_S,V_S\right] =O$ such that the interaction commutes with the free
Hamiltonian $H_S\otimes I_E.$ The following model has still $V=O$, but $%
\left[ H_S,V_S\right]$ does no longer vanish.

The Hilbert space of the model is ${\cal H}_{S+E}={\cal H}_S\otimes {\cal H}%
_E$ with ${\cal H}_S={\bf C}^2$ and ${\cal H}_E={\cal L}^2({\bf R}).$ The
Hamiltonian has the form (\ref{14a}) with $V=O$ and the following
specifications

$H_S\psi =(\overrightarrow{a}\overrightarrow{\sigma })\psi ,$ with $%
\overrightarrow{a}\in {\bf R}^3,$ $\overrightarrow{\sigma }$ Pauli matrices, 
$\psi \in {\bf C}^2,$

$H_Ef(x)=bx^2f(x)$ with a positive constant $b>0$, $f(x)\in {\cal L}^2({\bf R%
})$,

$V_S=\lambda \sigma _3,$ with a real coupling parameter $\lambda,$

$V_Ef(x)=xf(x)$, $f(x)\in {\cal L}^2({\bf R}).$

The statistical operator of the spin-$\frac 12$ system is a spin density
matrix (\ref{8}) with a polarization vector $\overrightarrow{p}\in B^3$. For
the total system we assume an initial state $W=\widehat{\rho }\left( 
\overrightarrow{p}\right) \otimes \omega $ where $\omega $ is a statistical
operator on ${\cal L}^2({\bf R})$ with a smooth integral kernel $\omega
(x,y).$ The time evolution $U(t)=\exp (-iHt)$ with the total Hamiltonian (%
\ref{14a}) then leads to the state $U(t)WU^{+}(t)$. The diagonal part of the
integral kernel of this statistical operator is $\left( U(t)WU^{+}(t)\right)
\left( x,x\right) =\omega (x,x)\widehat{\rho }\left( R_x(t)\overrightarrow{p}%
\right).$ Here $R_x(t)$ is the rotation induced by the $SU(2)$ matrix part
of $U(t)$, i.e. $\exp (-iht)$, with $h=(\overrightarrow{a}\overrightarrow{%
\sigma})+\lambda x\sigma_3.$ The reduced statistical operator $\rho (t)$ of
the spin-$\frac 12$ system is calculated as $\rho (t)={\rm tr}%
_EU(t)WU^{+}(t)=\int dx\omega (x,x)\widehat{\rho }\left( R_x(t)%
\overrightarrow{p}\right)$. For the initial state $\rho ={\rm tr}_EW=%
\widehat{\rho }\left( \overrightarrow{p}\right) $ the reduced dynamics then
leads to a statistical operator $\rho (t),$ which asymptotically approaches $%
\widehat{\rho }\left( \overrightarrow{q}\right) $, where the polarization
vector $\overrightarrow{q}$ is given by the linear mapping 
\begin{equation}
\label{18}\overrightarrow{q}=M\overrightarrow{p}:=\int dx\omega (x,x)%
\overrightarrow{n}(x)\left( \overrightarrow{p}\overrightarrow{n}(x)\right) 
\end{equation}
with the axis $\overrightarrow{n}(x)$ of the rotation $R_x(t)$. Under
appropriate conditions for the initial state $\omega $ of the environment,
the difference $\rho (t)-\widehat{\rho }\left( \overrightarrow{q}\right) $
can be uniformly estimated by
\noindent $\left\| \rho (t)-\widehat{\rho }\left( \overrightarrow{q}\right)
\right\| _1\leq c(1+\left| t\right| )^{-\gamma }.$ The mapping (\ref{18}) is
a symmetric contraction on ${\bf R}^3.$ We can distinguish two cases:

\smallskip
\noindent1) If $\overrightarrow{a}\parallel \overrightarrow{e_3}$ the
mapping (\ref{18}) reduces to $M\overrightarrow{p}=\overrightarrow{e_3}%
\left( \overrightarrow{e_3}\overrightarrow{p}\right) $, and we obtain the
results discussed in Sect.\ref{mod1}. The condition $\left| M\overrightarrow{%
p}\right| =\left| \overrightarrow{p}\right| $ is satisfied for $%
\overrightarrow{p}\parallel \overrightarrow{e_3},$ and only in this case $%
\rho (t)$ is not affected by the decoherence.

\noindent2) If $\overrightarrow{a}$ has components orthogonal to $%
\overrightarrow{e_3}$, also the direction of $M\overrightarrow{p}$ depends
on $\overrightarrow{p},$ and $\left| M\overrightarrow{p}\right| <\left| 
\overrightarrow{p}\right| $ holds for all vectors $\overrightarrow{p}\neq 
\overrightarrow{0}.$

\smallskip
\noindent In the second case there are no projection operators which commute
with all operators $\widehat{\rho }\left( M\overrightarrow{p}\right)$, $%
\overrightarrow{p}\in B^3.$ A superselection rule of the type (\ref{12}) can
hold in some approximative sense only if $a_1^2+a_2^2\ll a_3^2$. In that
case $M\overrightarrow{p}$ has very small components orthogonal to $%
\overrightarrow{e_3}$, and the off-diagonal matrix elements of the operators 
$\widehat{\rho }\left( M\overrightarrow{p}\right)$, $\overrightarrow{p}\in
B^3,$ are negligible.

\subsection{Models with scattering\label{mod3}}

We assume again a Hamiltonian (\ref{14a}) with $\left[H_S,V_S\right]=O $ as
in the Araki-Zurek model, but now with an additional scattering potential $V$
defined on the full Hilbert space ${\cal H}_{S+E}={\cal H}_S\otimes{\cal H}%
_E.$ Under appropriate conditions on $V$ the wave operator exists as strong
limit $\Omega =\lim _{t\rightarrow \infty }e^{iHt}e^{-iH_0t}$ on ${\cal H}%
_{S+E},$ where $H_0$ is the Hamiltonian of the Araki-Zurek model, $%
H_0=H_S\otimes I_E+I_S\otimes H_E+V_S\otimes V_E.$ (We assume for simplicity
that there are no bound states such that $\Omega ^{+}=\Omega ^{-1}.$) Then
the time evolution $U(t)=\exp (-iHt)$ behaves asymptotically as $%
U_0(t)\Omega^{+}$ with $U_0(t)=\exp (-iH_0t).$ More precisely we have for
all $W\in {\cal D(H}_{S+E}{\cal )}$%
\begin{equation}
\label{19}\lim _{t\rightarrow \infty }\left\| U(t)WU^{+}(t)-U_0(t)\Omega
^{+}W\Omega U_0^{+}(t)\right\| _1=0 
\end{equation}
in the trace norm. For initial states $W=\rho(0)\otimes \omega$ with smooth $%
\omega$, and for sufficiently regular scattering potentials the statistical
operator $\Omega^{+}W\Omega $ has only smooth contributions in ${\cal T}(%
{\cal H})$, and the reduced trace ${\rm tr}_EU_0(t)\Omega ^{+}W\Omega
U_0^{+}(t)$ yields the induced superselection sectors $P_m{\cal H}_S$ of
Sect.\ref{mod1}. But then (\ref{19}) implies for $\rho (t)={\rm tr}%
_EU(t)WU^{+}(t)$ the asymptotics $\lim _{t\rightarrow \infty }\left\| \rho
(t)-\sum_mP_m\rho (t)P_m\right\| _2=0.$ Hence $\rho(t)$ has again the
induced superselection sectors which originate from the spectrum (\ref{15})
of $V_S$. But in contrast to (\ref{13}) this bound is not uniform in the
initial state $\rho(0)$, since scattering does not allow a uniform estimate
for (\ref{19}).

\subsection{Concluding remarks}

The investigation of the models proves that the uniform emergence (\ref{13})
of superselection sectors is consistent with the mathematical rules of
quantum mechanics. But this result depends on rather restrictive assumptions
on the Hamiltonian. For more realistic models we can nevertheless expect a
strong quantitative suppression of the off--diagonal elements of the
statistical operator, as already proposed in [8].

\bigskip

\noindent{\large {\bf Acknowledgement}}

\medskip

\noindent I would like to thank S. M. Roy and the Theoretical Physics Group
of the Tata Institute for Fundamental Research for their kind hospitality.

\bigskip

\noindent{\large {\bf References}}

\medskip

\noindent[1] A~Einstein, B~Podolsky, and N~Rosen, {Phys. Rev.} {\bf 47},
777--780 (1935).

\noindent[2] E~Schr\"odinger, {Naturwiss.} {\bf 23}, 807--812, 823--828,
844--849 (1935).

\noindent[3] J~S Bell, {\em Speakable and unspeakable in quantum mechanics}.
(CUP, Cambridge, 1987).

\noindent[4] J~A Wheeler and W~H Zurek, {\em Quantum theory and measurement}
(Princeton University Press, 1983).

\noindent[5] N~N Bogolubov, A~A Logunov, A~I Oksak, and I~T Todorov, {\em {%
General principles of quantum field theory}}. (Kluwer, Dortrecht, 1990).

\noindent[6] A~S Wightman, {Nuovo Cimento} {\bf 110B}, 751 (1995)

\noindent[7] H~Araki, Prog. Theor. Phys. {\bf 64}, 719--730 (1980).

\noindent[8] W~H Zurek, {Phys. Rev.} {\bf D26}, 1862--1880 (1982).

\noindent[9] E~Joos and H~D Zeh, {Z. Phys.} {\bf B59}, 223--243 (1985).

\noindent[10] D~Giulini, E~Joos, C~Kiefer, J~Kupsch, I~O Stamatescu, and H~D
Zeh, {\em {Decoherence and the appearance of a classical world in quantum
theory}}.(Springer, Berlin, 1996).

\noindent[11] G Choquet, {\em {Lectures on analysis. Vol. II}} (New York,
Benjamin, 1969)

\noindent[12] E~M Alfsen, {\em Compact convex sets and boundary integrals},
(Springer, Berlin, 1971).

\noindent[13] E~C Beltrametti and G~Casinelli, {\em The logic of quantum
mechanics} (Reading, Addison-Wesley, 1981)

\noindent[14] R~Mirman, {Found. Phys.} {\bf 9}, 283--299 (1979).

\end{document}